\documentclass[12pt]{article}

\usepackage{amsmath}
\numberwithin{equation}{section}

\evensidemargin \oddsidemargin

\begin{document}

\makeatletter

\renewcommand{\thefootnote}{\fnsymbol{footnote}}
\newcommand{\beq}{\begin{equation}}
\newcommand{\eeq}{\end{equation}}
\newcommand{\bea}{\begin{eqnarray}}
\newcommand{\eea}{\end{eqnarray}}
\newcommand{\nn}{\nonumber\\}
\newcommand{\hs}[1]{\hspace{#1}}
\newcommand{\vs}[1]{\vspace{#1}}
\newcommand{\Half}{\frac{1}{2}}
\newcommand{\p}{\partial}
\newcommand{\ol}{\overline}
\newcommand{\wt}[1]{\widetilde{#1}}
\newcommand{\ap}{\alpha'}
\newcommand{\bra}[1]{\left\langle  #1 \right\vert }
\newcommand{\ket}[1]{\left\vert #1 \right\rangle }
\newcommand{\vev}[1]{\left\langle  #1 \right\rangle }

\newcommand{\ul}[1]{\underline{#1}}
\newcommand{\tr}{\mbox{tr}}
\newcommand{\ishibashi}[1]{\left\vert #1 \right\rangle\rangle }

\makeatother

\begin{titlepage}

\renewcommand{\thefootnote}{\fnsymbol{footnote}}

\hfill{hep-th/0306282}

\vspace{15mm}
\baselineskip 9mm
\begin{center}
{\LARGE \bf Boundary States\\
in IIA Plane-Wave Background}
\end{center}

\baselineskip 6mm
\vspace{10mm}
\begin{center}
Yeonjung Kim,$^a$\footnote{\tt geni@muon.kaist.ac.kr} 
and Jaemo Park$^b$\footnote{\tt jaemo@physics.postech.ac.kr} 
\\[5mm] 
{\sl $^a$Department of Physics, KAIST, 
Taejon 305-701, Korea \\ 
$^b$ Department of Physics, POSTECH, Pohang 790-784,
Korea \\} 
\end{center}

\vspace{20mm}

\thispagestyle{empty}

\vfill
\begin{center}
{\bf Abstract}
\end{center}
\noindent
We work out boundary states for Type IIA string theory on a plane wave background. 
By directly utilizing the channel duality, the induced conditions from the open 
string boundary conditions are imposed on the boundary states. The resulting 
boundary states correctly reproduce the partition functions of the open string 
theory for $DpDp$ and $DpD\bar{p}$ cases where $Dp$ branes are half BPS 
brane if located at the origin of the plane wave background.

\vspace{20mm}
\end{titlepage}

\baselineskip 6.5mm
\renewcommand{\thefootnote}{\arabic{footnote}}
\setcounter{footnote}{0}

\section{Introduction}
Recently there have been great interests on the string theory on the plane 
wave background\cite{bfhp}. Initially Metsaev worked out Type IIB string theory on 
the plane wave background in the lightcone gauge\cite{m, mt}. Subsequently the Type IIB 
string theory on the plane wave background attracted great interests in relation 
to the correspondence with N=4 Supersymmetric Yang-Mills theories\cite{bmn}. 

In a series of paper\cite{hyu074, sug029, hyu158, hyu090, hyunnew}, 
simple Type IIA string theory on the plane wave background 
have been studied. The background for the string theory is obtained by 
compactifying the 11-dimensional plane wave background on a circle and taking 
the small radius limit. The resulting string theory has many nice features.
It admits light cone gauge where the string theory spectrum is that of the free massive 
theory as happens in Type IIB theory. Furthermore the worldsheet enjoys (4,4) worldsheet 
supersymmetry. The structure of supersymmetry is simpler in the sense that the supersymmetry 
commutes with the Hamiltonian so that all members of the same supermultiplet
has the same mass. The various 1/2 BPS D-brane states were analyzed in the lightcone gauge, 
which gives consistent results with the BPS branes in the matrix model.
Subsequently, covariant analysis was carried out for the D-brane spectrums, which 
agrees with the previous results\cite{hyun123}. 

In this letter, we initiated the study of boundary states of Type IIA string theories. 
Partly we were motivated by the works in Type IIB side. In \cite{gaber}, the boundary 
states for Type IIB theories were worked out and were shown to exhibit interesting 
channel duality between open string and closed string. In \cite{gaber2, gaber3}, more 
general cases were studied. In their developments, the lightcone supersymmetries, 
kinematical as well as dynamical, have played a crucial role in the construction 
of the boundary states. However in the construction of the boundary states in the 
more general conformal field theories\cite{rec,rec2}, the supersymmetries are not important. 
And many of the theories considered do not have the spacetime supersymmetries. 
Rather, by directly utilizing the channel duality, one can first construct the boundary 
states, and then figure out the open string spectrums after the modular 
transformation and vice versa. In the simplest example of the flat space, starting from the 
open string boundary conditions, one can obtain the conditions to be imposed 
on the boundary states by simply interchanging the role of $\sigma$ and $\tau$
of the worldsheet coordinates\cite{pol, cal}. 

Indeed Michishita worked out Type IIB boundary 
states along this line of idea and obtained the consistent results with the previous 
works whenever the comparison is available\cite{yoji}. Also this approach was 
adopted in \cite{sken}. Here we adopt the same philosophy 
to obtain the boundary states in Type IIA string theory on the plane wave 
background. The resulting boundary states give the correct open string partition 
function after the modular transformation. 
In this letter we just work out the boundary states for $DpDp$ and 
$DpD\bar{p}$ case where $Dp$ branes are half BPS at the origin of the plane wave.
Obviously there are more general cases to be studied. The detailed explorations 
of the boundary states of the IIA theory will appear elsewhere\cite{park}. 
An interesting fact for these $Dp$ branes is that they do not have the dynamical 
supersymmetries away from the origin of the plane wave background \cite{hyun123}.
Similar phenomenon occurs in Type IIB cases as well \cite{sken2, peeters}.  
As we finish the draft, we are aware of the work\cite{shin} where Type IIA boundary 
states are constructed following the approach of \cite{gaber, gaber2}. 

\section{Free string theory and D-branes in IIA plane-wave background}
We closely follow the notation of \cite{hyu158} to describe the Type IIA string theory on the 
plane wave background. The background is given by 
\beq
ds^2 = -2dX^+dX^--A(x^I)(dX^+)^2+\sum_{i=1}^{8}dX^idX^i,
\eeq
\beq
F_{+123} =\mu , F_{+4} =\frac{\mu}{3}
\eeq
and 
\beq
A(x^I)=\sum_{i=1}^{4}\frac{\mu^2}{9}(X^{i})^2+\sum_{i'=5}^{8}\frac{\mu^2}{36}(X^{i'})^2
\eeq
where we define$X^\pm=\frac{1}{\sqrt{2}}(X^0\pm X^9)$. Throughout this letter, 
we use the convention 
that 
unprimed coordinates denote 1,2,3,4 directions 
while primed coordinates denote 5,6,7,8 directions. 
Also 
unprimed quantities are associated with 1,2,3,4 directions and primed ones are related 
to 5,6,7,8 directions. 
For the worldsheet coordinates, we use $\p_\pm=\Half(\p_\tau\pm\p_\sigma)$.
The worldsheet action for the closed string is given by 
\bea
S_{LC} & = &- \frac{1}{4\pi\ap}\int d^2\sigma (-\p_\mu X^+\p^\mu X^-
+\frac{1}{2}\p_\mu X^i\p^\mu X^i + \frac{m^2}{9}\sum_{i=1}^{4}X^iX^i
+\frac{m^2}{36}\sum_{i'=1}^{4}X^{i'}X^{i'}
\nn
 & & \sum_{b=\pm} (-i\psi_b^1\p_{+}\psi_b^1-i\psi_b^2\p_{-}\psi_b^2)
+\frac{2im}{3}\psi_{+}^2\gamma^4\psi_{-}^1
-\frac{im}{3}\psi_{-}^2\gamma^4\psi_{+}^1)
\eea 
where $m=\ap p^+\mu$ and 
$\gamma^i$ are $8\times 8$ matrices satisfying $\{\gamma^i, \gamma^j\}=\delta^{ij}$. 
The sign of subscript $\psi^A_{\pm}$ denotes the eigenvalue of $\gamma^{1234}$ while the 
superscript $A=1,2$ denotes the eigenvalue of $\gamma^9$. The theory of interest contains two 
supermultiplets $(X^i, \psi_-^1, \psi_+^2)$ and $(X^{i'}, \psi_+^1, \psi_-^2)$ 
of (4,4) worldsheet supersymmetry with the masses $\frac{m}{3}$ and  $\frac{m}{6}$ respectively. 
The fermions of the first supermultiplet has $\gamma^{12349}$ eigenvalue of 1 while those 
of the second has the eigenvalue of $-1$.

The mode expansion for the bosonic coordinates is given by 
\bea
X^i & = & i\sqrt{\frac{\ap}{2}}
\sqrt{\frac{1}{\omega_0}}
(a^i_0e^{-i\omega_0\tau}-a_0^{i\dagger} e^{i\omega_0\tau})
+i\sqrt{\frac{\ap}{2}}\sum_{n=1}^\infty\frac{1}{\sqrt{\omega_n}}\{
e^{-i\omega_n\tau}
(a^i_ne^{in\sigma}+\wt{a}_n^ie^{-in\sigma})
\nn
& & -e^{i\omega_n\tau}
(a_n^{i\dagger}e^{-in\sigma}+\wt{a}_n^{i\dagger}e^{in\sigma})\}
\eea
where we have $\omega_n=\sqrt{(\frac{m}{3})^2+n^2}$ with $n\geq 0$ and 
\beq
a_0^i\equiv\frac{\sqrt{\frac{\ap}{\omega_0}}p^i-i\sqrt{\frac{\omega_0}{\ap}}x^i}{\sqrt{2}},
\eeq
\beq
a_0^{i\dagger}\equiv\frac{\sqrt{\frac{\ap}{\omega_0}}p^i
+i\sqrt{\frac{\omega_0}{\ap}}x^i}{\sqrt{2}}.   \nonumber
\eeq
The commutation relation is given by
\bea
& & \quad [a^i_n,a^{j\dagger}_{m}]=\delta^{ij}\delta_{nm}, \,\, n,m\geq 0   \\
& & \quad [\wt{a}^i_n,\wt{a}^{j\dagger}_{m}]=\delta^{ij}\delta_{nm},  \,\,  n,m > 0. \nonumber
\eea
And the mode expansion for the primed coordinates is given by 
\bea
X^{i'} & = & i\sqrt{\frac{\ap}{2}}
\sqrt{\frac{1}{\omega_0'}}
(a^{i'}_0e^{-i\omega_0'\tau}-a_0^{i'\dagger} e^{i\omega_0'\tau})
+i\sqrt{\frac{\ap}{2}}\sum_{n=1}^\infty\frac{1}{\sqrt{\omega_n'}}\{
e^{-i\omega_n'\tau}
(a^{i'}_ne^{in\sigma}+\wt{a}^{i'}_ne^{-in\sigma})
\nn
& & -e^{i\omega_n'\tau}
(a_n^{i'\dagger}e^{-in\sigma}+\wt{a}_n^{i'\dagger}e^{in\sigma})\}
\eea
where we have  $\omega_n'=\sqrt{(\frac{m}{6})^2+n^2}$ with $n\geq 0$ and 
\beq
a_0^{i'}\equiv\frac{\sqrt{\frac{\ap}{\omega_0'}}p^{i'}
-i\sqrt{\frac{\omega_0}{\ap}}x^{i'}}{\sqrt{2}},
\eeq
\beq
a_0^{i\dagger'}\equiv\frac{\sqrt{\frac{\ap}{\omega_0'}}p^{i'}
+i\sqrt{\frac{\omega_0'}{\ap}}x^{i'}}{\sqrt{2}}.   \nonumber
\eeq
The commutation relation is given by
\bea
& & \quad [a^{i'}_n,a^{j'\dagger}_{m}]=\delta^{i'j'}\delta_{nm}, \,\,\, n,m\geq 0  \\
& & \quad [\wt{a}^{i'}_n,\wt{a}^{j'\dagger}_{m}]=\delta^{i'j'}\delta_{nm}, \,\,\, n,m > 0. \nonumber
\eea
The bosonic part of the lightcone Hamiltonian $H_{bc}$ is 
\bea
H_{bc} & = & \frac{1}{2\pi\ap p^+}\int_0^{2\pi} d\sigma_0(\Half P^iP^i
+\frac{1}{2} \p_\sigma X^i\p_\sigma X^i+\Half (\frac{m}{3})^2X^iX^i \nonumber \\
  & &+\Half P^{i'}P^{i'}
+\frac{1}{2} \p_\sigma X^{i'}\p_\sigma X^{i'}+\Half (\frac{m}{6})^2X^{i'}X^{i'})
\eea
with $P^i=\p_{\tau}X^i$ and $P^{i'}=\p_{\tau}X^{i'}$. In terms of the oscillators 
\beq
p^+ H_{bc}= \omega_0a_0^{i\dagger}a^i_0+\omega_0'a_0'^{i\dagger}a^{i'}_0
+\sum_{n=1}^\infty\omega_n(a_n^{i\dagger} a^i_n
 +\wt{a}_n^{i\dagger}\wt{a}^i_n)
+\omega_n'(a_n^{i\dagger'} a^{i'}_n
 +\wt{a}_n^{i\dagger'}\wt{a}^{i'}_n)+e_{bc0}   \label{hb}
\eeq
where $e_{bc0}$ is the zero-point energy contribution. Throughout the letter,
repeated indices $i$ and $i'$
are assumed to be summed over 1,2,3,4 and 5,6,7,8 respectively.  
The fermionic mode expansion is given by
\bea
\psi_-^1 & = & \sqrt{\frac{\ap}{2}}
(\chi e^{-i\omega_0\tau}+\chi^\dagger e^{i\omega_0\tau})
+\sum_{n=1}^\infty \sqrt{\ap}c_n\{
e^{-i\omega_n\tau}
(\wt{\psi}_ne^{in\sigma}-i\frac{\omega_n-n}{\omega_0}\gamma^4\psi_ne^{-in\sigma})
\nn
& & +e^{i\omega_n\tau}
(\wt{\psi}_n^\dagger e^{-in\sigma}
+i\frac{\omega_n-n}{\omega_0}\gamma^4\psi_n^\dagger e^{in\sigma})\} \label{f1}
\eea
where $c_n=1/\sqrt{1+(\frac{\omega_n-n}{\omega_0})^2}$ and 
\beq
\chi=\frac{\wt{\psi_0}-i\gamma^4\psi_0}{\sqrt{2}},  \,\,\, 
\chi^\dagger=\frac{\wt{\psi_0}+i\gamma^4\psi_0}{\sqrt{2}}
\eeq
with
\bea
\psi_+^2 & = & \sqrt{\frac{\ap}{2}}
(i\gamma^4\chi e^{-i\omega_0\tau}-i\gamma^4\chi^\dagger e^{i\omega_0\tau})
+\sum_{n=1}^\infty \sqrt{\ap}c_n\{
e^{-i\omega_n\tau}
(\psi_ne^{-in\sigma}+i\frac{\omega_n-n}{\omega_0}\gamma^4\wt{\psi}_ne^{in\sigma})
\nn
& & +e^{i\omega_n\tau}
(\psi_n^\dagger e^{in\sigma}
-i\frac{\omega_n-n}{\omega_0}\gamma^4\wt{\psi}_n^\dagger e^{-in\sigma})\}. \label{f2}
\eea
The anticommutation relations are 
\bea
& & \quad \{\psi^a_n,\psi^{b\dagger}_{m}\}=\delta^{ab}\delta_{nm}, \,\,\, n,m >0 \nonumber \\
& & \quad \{\wt{\psi}^a_n,\wt{\psi}^{b\dagger}_{m}\}=\delta^{ab}\delta_{nm},  
\,\,\, n,m >0 \nonumber \\ 
& & \quad \{\chi^a, \chi^{b\dagger}\}=\delta^{ab} 
\eea
where $a,b$ range from 1 to 4. 
The mode expansion of the superpartners for $X^{i'}$ is 
\bea
\psi_+^1 & = & \sqrt{\frac{\ap}{2}}
(\chi'e^{-i\omega_0'\tau}+\chi'^\dagger e^{i\omega_0'\tau})
+\sum_{n=1}^\infty \sqrt{\ap}c_n'\{
e^{-i\omega_n'\tau}
(\wt{\psi'}_ne^{in\sigma}+i\frac{\omega_n'-n}{\omega_0'}\gamma^4\psi_n'e^{-in\sigma})
\nn
& & +e^{i\omega_n'\tau}
(\wt{\psi}_n'^\dagger e^{-in\sigma}
-i\frac{\omega_n'-n}{\omega_0'}\gamma^4\psi_n'^\dagger e^{in\sigma})\}  \label{f3}
\eea
where $c_n'=1/\sqrt{1+(\frac{\omega_n'-n}{\omega_0'})^2}$ and
\beq
\chi'=\frac{\wt{\psi_0'}+i\gamma^4\psi_0'}{\sqrt{2}},   \,\,\,
\chi^{'\dagger}=\frac{\wt{\psi_0'}-i\gamma^4\psi_0'}{\sqrt{2}}
\eeq
with
\bea
\psi_-^2 & = & \sqrt{\frac{\ap}{2}}
(-i\gamma^4\chi'e^{-i\omega_0'\tau}+i\gamma^4\chi'^\dagger e^{i\omega_0'\tau})
+\sum_{n=1}^\infty \sqrt{\ap}c_n'\{
e^{-i\omega_n'\tau}
(\psi_n'e^{-in\sigma}-i\frac{\omega_n'-n}{\omega_0'}\gamma^4\wt{\psi}_n'e^{in\sigma})
\nn
& & +e^{i\omega_n'\tau}
(\psi_n'^\dagger e^{in\sigma}
+i\frac{\omega_n'-n}{\omega_0'}\gamma^4\wt{\psi}_n'^\dagger e^{-in\sigma})\}. \label{f4}
\eea
The anticommutation relation is given by
\bea
& & \quad \{\psi^{'a}_n,\psi^{'b\dagger}_{m}\}=\delta^{ab}\delta_{nm}, \,\,\, n,m>0 \nonumber \\
& & \quad \{\wt{\psi}^{'a}_n,\wt{\psi}^{'b\dagger}_{m}\}=\delta^{ab}\delta_{nm}, 
\,\,\, n,m>0 \nonumber \\
& & \quad \{\chi^{'a}, \chi^{'b\dagger}\}=\delta^{ab}. 
\eea
The fermionic contribution to the Hamiltonian is 
\bea
p^+H_{fc}
& = & \omega_0\chi^\dagger \chi+\omega_0'\chi'^\dagger \chi'
+\sum_{n=1}^\infty\omega_n(\psi_n^\dagger \psi_n+\wt{\psi}_n^\dagger \wt{\psi}_n)
+\omega_n(\psi_n^\dagger \psi_n+\wt{\psi}_n^\dagger \wt{\psi}_n)+e_{fc0} . \label{hf}
\eea
For the closed string, the zero point energy contribution is 
zero, i.e., $e_{bc0}+e_{fc0}=0$.

Now we discuss the Hamiltonian of the open strings and their mode expansions.
First we deal with the bosonic part of the Hamiltonian
\bea
H_b & = & \frac{1}{2\pi\ap p^+}\int_0^{\pi} d\sigma(\Half P^iP^i
+\Half \p_\sigma X^i\p_\sigma X^i+\omega_0^2X^iX^i \nn
 &  &+\Half P^{i'}P^{i'} +\Half \p_\sigma X^{i'}\p_\sigma X^{i'}+\omega_0^{'2}X^{i'}X^{i'})
\eea
with $P^i=\p_\tau X^i$ and $P^{i'}=\p_\tau X^{i'}$.
For the Neumann boundary condition, 
\bea
X^{i_N}&=&i\sqrt{\frac{\ap}{\omega_0}}
(a^{i_N}_0e^{-i\omega_0\tau}-a_0^{i_N\dagger} e^{i\omega_0\tau})  \nonumber \\
&+& i\sqrt{\frac{\ap}{2}} \sum_{n=1}^\infty\frac{1}{\sqrt{\omega_n}}
(a^{i_N}_ne^{-i\omega_n\tau}-a_n^{i_N\dagger} e^{i\omega_n\tau})
(e^{in\sigma}+e^{-in\sigma}),
\eea
with the commutation relation being
\beq
[a_n^{i_N}, a_m^{j_N\dagger}]=\delta_{nm}\delta^{i_Nj_N},  \,\,\,\, i_N,j_N\geq 0.
\eeq
The contribution to the Hamiltonian is 
\beq
p^+H_{b1}=\sum_{n=1}^\infty\omega_n
a^{i_N}_na_n^{i_N\dagger}+e_{b01N}.
\eeq
For the Dirichlet boundary conditions with $x^{i_D}(\sigma=0)=x^{i_D}_0,  
x^{i_D}(\sigma=\pi)=x^{i_D}_1$
\bea
X^{i_D} & = & \frac{x_0^{i_D}(e^{\omega_0\sigma}-e^{-\omega_0\sigma})-
 x_0^{i_D}(e^{\omega_0(\sigma-\pi)}-e^{-\omega_0(\sigma-\pi)})}{e^{\pi\omega_0}-e^{-\pi\omega_0}}
\nn
 & & +i\sqrt{\frac{\ap}{2}}\sum_{n=1}^\infty\frac{1}{\sqrt{\omega_n}}
(a_n^{i_D}e^{-i\omega_n\tau}+a_n^{i_D\dagger} e^{i\omega_n\tau})
(e^{in\sigma}-e^{-in\sigma})
\eea
with the commutation relation 
\beq
[a_n^{i_D}, a_m^{j_D\dagger}]=\delta_{nm}\delta^{i_Dj_D},  \,\,\,\, i_D,j_D > 0.
\eeq
The contribution to the Hamiltonian is 
\bea
p^+ H=\frac{\omega_0}{4\pi\ap}\frac{(e^{\pi\omega_0}+e^{-\pi\omega_0})
((x_1^{i_D})^2+(x_0^{i_D})^2-4x_1^{i_D}x_0^{i_D})}{e^{\pi\omega_0}-e^{-\pi\omega_0}}
+\sum_{n=1}^\infty\omega_n
a^{i_D}_na_n^{i_D\dagger} +e_{boD}.
\eea
For the $x^{i'}$ directions, we have the similar expressions, but we should 
replace unprimed quantities by primed quantities, e.g. $\omega_n$ by $\omega_n'$. 
The total bosonic Hamiltonian is given by
\bea
p^+ H_{bo} &=& \sum_{n=1}^\infty(\omega_na^{i_D}_na_n^{i_D\dagger}
+\omega_na^{i'_D}_na_n^{i'_D\dagger}
+\omega_na^{i_N}_na_n^{i_N\dagger}+\omega_na^{i'_N}_na_n^{i'_N\dagger})  \nonumber \\
& &+\frac{\omega_0}{4\pi\ap}\frac{(e^{\pi\omega_0}+e^{-\pi\omega_0})
((\vec{x_1^D})^2+(\vec{x_0^D})^2-4\vec{x_1^D}\cdot\vec{x_0^D})}{e^{\pi\omega_0}-e^{-\pi\omega_0}}
\nonumber \\
& &+\frac{\omega'_0}{4\pi\ap}\frac{(e^{\pi\omega'_0}+e^{-\pi\omega'_0})
((\vec{x_1^{'D}})^2+(\vec{x_0^{'D}})^2-4\vec{x_1^{'D}}\cdot\vec{x_0^{'D}})}
{e^{\pi\omega_0'}-e^{-\pi\omega_0'}} +e_{b0}.  \label{hbo}
\eea

For the fermion, we impose the boundary conditions
\beq
\psi^1_{\pm} |_{\sigma=0, \pi}=\eta \Omega \psi^2_{\mp} |_{\sigma=0, \pi}.
\eeq
Here $\eta=1$ is $DpDp$ case, which means that open string ends on a $Dp$ brane at one end 
and ends on another $Dp$ brane at the other end. And $\eta=-1$ denotes $DpD\bar{p}$ case where $D\bar{p}$ 
means an anti Dp brane. In this letter we only deal with $DpDp$ and $DpD\bar{p}$ cases, but 
certainly more general possibilities exist, which will be explored elsewhere. 
Let us discuss $\eta=+1$ case first. 
In order for the Dp brane to have the supersymmetry at the origin of the 
plane wave background, $\Omega$ should satisfy 
\bea
\gamma^4\Omega\gamma^4\Omega & = & -1 \nonumber \\
\{\Omega, \gamma^9\} & = & 0 \nonumber  \\
\{\Omega, \gamma^{1234}\} & = & 0.
\eea

The actual form of the $\Omega$ depends on the dimension of the world volume. 
D2, D4, D6, D8 branes were shown to have the supersymmetry at the origin and the 
expressions of $\Omega$ are tabulated at \cite{hyu158}. However the detailed form of $\Omega$ 
is not important for the subsequent discussions. 
One interesting point is that depending on the position, the number of the 
supersymmetries to be preserved by the D-brane is different. The D-branes away from the 
origin have no dynamical supersymmetry as shown in the covariant analysis\cite{hyun123}. 
From the conditions above, one obtain the conditions for the open string modes
\beq
\wt{\psi}_n=\Omega \psi_n,  \wt{\psi}'_n=\Omega \psi'_n, 
\eeq
for all $n$. To obtain the open string mode expansion one should replace 
$\wt{\psi}_n$, $\wt{\psi}'_n$ by the above conditions in the closed string expressions 
(\ref{f1}), (\ref{f2}), (\ref{f3}) and (\ref{f4}). 
The fermionic contribution to the Hamiltonian is
\bea
p^+H_{fo} &=& \sum_{n=1}^\infty(\omega_n \psi_n^\dagger\psi_n
+\omega_n \psi_n^{'\dagger}\psi'_n) 
\nonumber \\
  & & +(\frac{i}{2}\omega_0\psi_0\gamma^4\Omega\psi_0
        -\frac{i}{2}\omega'_0\psi'_0\gamma^4\Omega\psi'_0)+e_{f0}.  \label{hfo}
\eea
For $\eta=1$, the zero point energy contribution for Dp brane is given by
$e_{b0}+e_{f0}=\frac{1}{2}\omega_0n_N+\frac{1}{2}\omega'_0n'_N$
with $n_N+n'_N=p-1$ where $n_N$ is the number of Neumann directions among 1,2,3,4 
directions and $n_{N'}$ is the number of Neumann directions among 5,6,7,8 
directions. 

For the $DpD\bar{p}$, we have $\eta=-1$ and the fermion modes have half-integer modings in
worldsheet coordinates $\sigma$ and the energy eigenvalues are given by 
$\bar{\omega}_n\equiv \sqrt{(\frac{m}{3})^2+(n-\frac{1}{2})^2}$, 
$\bar{\omega}'_n\equiv \sqrt{(\frac{m}{6})^2+(n-\frac{1}{2})^2}$ and 
\beq
p^+H_f = \sum_{n=1}^\infty(\bar{\omega}_n \psi_{n-\frac{1}{2}}^\dagger
\psi_{n-\frac{1}{2}}+\bar{\omega}'_n \psi_{n-\frac{1}{2}}^{'\dagger}
\psi'_{n-\frac{1}{2}})+e_{f0}'
\eeq
where $\{\psi^a_{n-\frac{1}{2}}, \psi^{b\dagger}_{n-\frac{1}{2}}\}=\delta^{ab}_{nm}$ 
and the similar relation holds for $\psi'$s. 
The value of zero point energy $e_{b0}+e_{f0}'$ will be given shortly after introducing suitable 
quantities.

Now  one can evaluate the cylinder amplitude for the open string
\beq
Z_{O}=\int_0^\infty\frac{dt}{t}\int\ap dp^+dp^-
Tr e^{-2\pi t(-2\ap p^+p^-+p^+H)}. 
\eeq
where $H=H_{bo}+H_{fo}$ of (\ref{hbo}) and (\ref{hfo}). 
For the evaluation, it is convenient to define 
\bea
f_1^{(m)}(q) & = & q^{-\Delta_m}(1-q^m)^{1/2}\prod_{n=1}^\infty
(1-q^{\sqrt{n^2+m^2}})  \nonumber  \\
f_2^{(m)}(q) & = & q^{-\Delta_m}(1+q^m)^{1/2}\prod_{n=1}^\infty
(1+q^{\sqrt{n^2+m^2}})  \nonumber  \\
f_3^{(m)}(q) & = & q^{-\Delta'_m}\prod_{n=1}^\infty
(1+q^{\sqrt{(n-\frac{1}{2})^2+m^2}})  \nonumber  \\
f_4^{(m)}(q) & = & q^{-\Delta'_m}\prod_{n=1}^\infty
(1-q^{\sqrt{(n-\frac{1}{2})^2+m^2}})
\eea
where $q=e^{-2\pi t}$ and $\Delta_m$, $\Delta'_m$ are defined as 
\bea 
\Delta_m &=& -\frac{1}{(2\pi)^2}\sum_{p=1}^\infty\int_0^\infty
e^{-p^2s}e^{-\frac{\pi^2m^2}{s}}
\nonumber \\
\Delta'_m &=& -\frac{1}{(2\pi)^2}\sum_{p=1}^\infty(-1)^p\int_0^\infty
e^{-p^2s}e^{-\frac{\pi^2m^2}{s}}
\eea
The modular transformation property is proved to be\cite{gaber} 
\beq
f_1^{(m)}(q)=f_1^{(\hat{m})}(\tilde{q}), f_2^{(m)}(q)=f_4^{(\hat{m})}(\tilde{q}), 
f_3^{(m)}(q)=f_3^{(\hat{m})}(\tilde{q})
\eeq
where $\hat{m}=mt$ and $\tilde{q}=e^{-\frac{2\pi}{t}}$
With this expression, the cylinder amplitude is easily evaluated following \cite{gaber}.
The result is 
\bea
Z_{O} & = & \int_0^\infty\frac{dt}{2 t^2}
(2{\rm sinh}\pi t \omega_0)^{2-n_N}(2{\rm sinh}\pi t \omega'_0)^{2-n'_N}
\frac{f_A^{\omega_0}(q)^4f_A^{\omega'_0}(q)^4}{f_1^{\omega_0}(q)^4f_1^{\omega'_0}(q)^4}
\nonumber \\
& & \exp{-2\pi t(\frac{\omega_0}{4\pi\ap}\frac{(e^{\pi\omega_0}+e^{-\pi\omega_0})
((\vec{x_1^D})^2+(\vec{x_0^D})^2)
-4\vec{x_1^D}\cdot\vec{x_0^D}}{e^{\pi\omega_0}-e^{-\pi\omega_0}})} \nonumber \\
& & \exp{-2\pi t(\frac{\omega_0'}{4\pi\ap}\frac{(e^{\pi\omega_0'}+e^{-\pi\omega_0'})
((\vec{x_1^{'D}})^2+(\vec{x_0^{'D}})^2)
-4\vec{x_1^{'D}}\cdot\vec{x_0^{'D}}}{e^{\pi\omega_0'}-e^{-\pi\omega_0'}})} \label{ans}
\eea
where A=1 for $DpDp$ and A=4 for $DpD\bar{p}$. The last two lines of (\ref{ans}) are coming from 
the Dirichlet directions. 
Using the result of \cite{gaber}, one can see that 
zero energy contribution for $DpD\bar{p}$ is 
$\frac{n_N-2}{2}\omega_0+4\Delta_{\omega_0}-4\Delta'_{\omega_0}
+\frac{n_{N'}-2}{2}\omega'_0+4\Delta_{\omega'_0}-4\Delta'_{\omega'_0}$ See also \cite{cot}. 
After the modular transformation, with $t=\frac{1}{2\ell}, \hat{\omega_0}=\omega_0 t$,
we have 
\bea
Z_{C_2} & = & \int_0^\infty d\ell
(2{\rm sinh}\pi \hat{\omega_0})^{2-n_N}(2{\rm sinh}\pi \hat{\omega'_0})^{2-n'_N}
\frac{f_B^{\hat{\omega_0}}(\tilde{q})^4f_B^{\hat{\omega'_0}}(\tilde{q})^4}
{f_1^{\hat{\omega_0}}(\tilde{q})^4 f_1^{\hat{\omega'_0}}(\tilde{q})^4}
\nonumber \\
& & \exp{-(\frac{\hat{\omega_0}}{2\ap}\frac{(e^{2\ell\hat{\omega_0}\pi}
+e^{-2\ell\hat{\omega_0}\pi})
((\vec{x_1^D})^2+(\vec{x_0^D})^2)-4\vec{x_1^D}\cdot\vec{x_0^D}}{e^{2\ell\hat{\omega_0}\pi}
-e^{-2\ell\hat{\omega_0}\pi}})} \nonumber \\
& & \exp{-(\frac{\hat{\omega_0'}}{2\ap}\frac{(e^{2\ell\hat{\omega_0'}\pi}
+e^{-2\ell\hat{\omega_0'}\pi})
((\vec{x_1^{'D}})^2+(\vec{x_0^{'D}})^2)-4\vec{x_1^{'D}}\cdot\vec{x_0^{'D}}}
{e^{2\ell\hat{\omega_0'}\pi}
-e^{-2\ell\hat{\omega_0'}\pi}})} 
\eea
with $\tilde{q}=e^{-4\pi\ell}$ where $B=1$ for $DpDp$ and $B=2$ for $DpD\bar{p}$.
This will be compared with the overlap of the boundary states in the next section.

\section{Boundary state construction}

As explained in  \cite{gaber}, if we have the usual lightcone gauge in the open string 
we have to take the nonstandard lightcone gauge in the closed string channel 
where the role of $X^+$ and $p^+$ are reversed. In this case, 
\beq
X^+H_c=p^+ H_{bc}+p^+ H_{fc}
\eeq
where $p^+H_{bc}$ and $p^+H_{fc}$ are given by (\ref{hb}) and (\ref{hf})  respectively
with replacing $m$ by $\hat{m}\equiv mt$. This change of the mass parameter is due to the 
conformal transformation needed going from the open string channel to the closed string 
channel \cite{gaber}.
If we have a channel duality between closed strings and open strings 
we can obtain the condition for the boundary states from the boundary conditions 
for the open strings by interchanging the role of $\sigma$ and $\tau$ coordinates of 
the world sheets.

For the Neumann boundary conditions for a bosonic coordinate 
$X$, $\partial_{\sigma}X=0$ at $\sigma=0$, 
the corresponding condition for the boundary states is $\partial_{\tau}X=0$ at $\tau=0$.
This gives the condition
\beq
a_n+\wt{a}_n^\dagger=0, \wt{a}_n+a_n^\dagger=0, 
    a_0+a_0^\dagger=0
\eeq
whose solution is 
\beq
\exp(-\sum_{n\geq 1}a_n^\dagger\wt{a}_n^\dagger
     -\Half(a_0^\dagger)^2)\ket{0}.
\eeq
Here $\ket{0}$ is the vacuum which is annihilated by $a_n, \wt{a}_n, a_0$. 
Its CPT conjugate state is given by 
\beq
\bra{0}\exp(-\sum_{n\geq 1}a_n\wt{a}_n-\Half a_0^2)
\eeq
For the Dirichlet conditions, we impose 
$\partial_{\sigma}X=0$ at $\tau=0$ with $x^i=x^i_0$.
We have 
\beq
  a_n-\wt{a}_n^\dagger=0, \wt{a}_n-a_n^\dagger=0 \,\, (n\geq 1),
   \,\,\,\, a_0-a_0^\dagger=-i\sqrt{\frac{2\omega_0}{\ap}}x_0
\eeq
whose solution is 
\beq
\exp(\Half(a_0^\dagger-i\sqrt{\frac{2\omega_0}{\ap}}x_0)^2
 +\sum_{n\geq 1}a_n^\dagger\wt{a}_n^\dagger)\ket{0}
\eeq
For the evaluation of the overlap of the boundary conditions we need 
\bea
& & \bra{0}\exp(\Half(a_0+i\sqrt{\frac{2\omega_0}{\ap}}x_1)^2)
e^{-2\pi\ell \omega_0a_0^\dagger a_0}
\exp(\Half(a_0^\dagger-i\sqrt{\frac{2\omega_0}{\ap}}x_0)^2)\ket{0}
\nn
& = & (1-e^{-4\pi\ell\omega_0})^{-1/2}
 \exp(-\frac{\omega_0}{\ap}\frac{x_0^2+x_1^2-2x_0x_1e^{-2\pi\ell\omega_0}}
 {1-e^{-4\pi\ell\omega_0}})
\nn
& = & \frac{\exp(-\frac{\omega_0}{2\ap}(x_0^2+x_1^2))}{(1-e^{-4\pi\ell\omega_0})^{1/2}}
 \exp(-\frac{\omega_0}{2\ap}\frac{(e^{2\pi\ell\omega_0}+
 e^{-2\pi\ell\omega_0})(x_0^2+x_1^2)-4x_0x_1}
 {e^{2\pi\ell\omega_0}-e^{-2\pi\ell\omega_0}})
\eea
Thus the total bosonic boundary state is given by 
\bea
& & \exp(-\sum_{n=1}^\infty a_n^{i_N\dagger}\wt{a}_n^{i_N\dagger}
     -\Half(a_0^{i_N\dagger})^2-a_n^{i'_N\dagger}\wt{a}_n^{i'_N\dagger}
     -\Half(a_0^{i'_N\dagger})^2) \nonumber  \\
& & \exp(\Half(a_0^{i_D\dagger}-i\sqrt{\frac{2\omega_0}{\ap}}x_0^{i_D})^2
 +\sum_{n= 1}^\infty a_n^{i_D\dagger}\wt{a}_n^{i_D\dagger})  \nonumber  \\
& & \exp(\Half(a_0^{i'_D\dagger}-i\sqrt{\frac{2\omega_0}{\ap}}x_0^{i'_D})^2
 +\sum_{n= 1}^\infty a_n^{i'_D\dagger}\wt{a}_n^{i'_D\dagger}) \ket{0}
\eea
where the bosonic vacuum $\ket{0}$ is annihilated by all lowering operators 
associated with the raising operators appearing in the expression.

Now the boundary condition for the fermions at the open string channel
\beq
\psi^1_{\pm}|_{\sigma=0}=\eta\psi^2_{\mp}|_{\sigma=0}
\eeq
is translated into the conditions at the closed string channel
\beq
\psi^1_{\pm}|_{\tau=0}=i\eta\psi^2_{\mp}|_{\tau=0}.
\eeq
From $\psi^1_- =i\eta\psi^2_{+}$ at $\tau=0$, we obtain
\beq 
\chi^\dagger=-\eta\Omega\gamma^4\chi, \,\,  \chi=\eta\Omega\gamma^4\chi^\dagger.
\eeq
which are equivalent if we use $\gamma^4\Omega\gamma^4\Omega
=\Omega\gamma^4\Omega\gamma^4=-1$,
which can be proved by using the fact that $\Omega$ and $\gamma^4$ 
are either commute or anticommute.
For nonzero modes we have 
\beq
\wt{\psi_n}=i\eta\Omega\psi_n^\dagger, \psi_n=-i\eta\Omega^T\wt{\psi}_n^\dagger
\eeq
which lead to the same boundary state. From the above conditions we obtain 
\beq
\exp(\frac{1}{2}\eta\chi^\dagger\Omega\gamma^4\chi^\dagger
+i\eta\sum_{n=1}^\infty\wt{\psi}_n^\dagger\Omega\psi_n^\dagger\ket{0}
\eeq
where the fermionic vacuum state $\ket{0}$ is annihilated by 
$\psi_n, \wt{\psi}_n$ and $\chi$. 
From $\psi^1_+ =i\eta\psi^2_{-}$ at $\tau=0$ gives 
\beq 
\chi'=-\eta\Omega\gamma^4\chi^{'\dagger},  \,\, \chi^{'\dagger}=\eta\Omega\gamma^4\chi'.
\eeq
and 
\beq
\wt{\psi_n'}=i\eta\Omega\psi_n^{'\dagger}, \,\, \psi_n'=-i\eta\Omega^T\wt{\psi}_n^{'\dagger}.
\eeq
The total fermionic boundary state is given by 
\beq
\exp(\frac{1}{2}\eta\chi^\dagger\Omega\gamma^4\chi^\dagger
-\frac{1}{2}\eta\chi^{'\dagger}\Omega\gamma^4\chi^{'\dagger}
+i\eta\sum_{n=1}^\infty\wt{\psi}_n^\dagger\Omega\psi_n^\dagger
+i\eta\sum_{n=1}^\infty\wt{\psi}_n^{'\dagger}\Omega\psi_n^{'\dagger}\ket{0}
\eeq
where the fermionic vacuum $\ket{0}$ is annihilated by all of the corresponding 
lowering operators for the raising operators 
appeared in the expressions. The total boundary states are the product of the bosonic 
boundary state and the fermionic one.  
Boundary states at other $\tau=\tau_0$ are generated by the closed string Hamiltonian
acting on boundary states at $\tau=0$, 
i.e, $\ket{B, \tau=\tau_0}=e^{iX^+H_c\tau_0}\ket{B, \tau=0}$. From now on  all boundary 
states are assumed to be those at $\tau=0$. 

Let $\ket{B_0, \eta}$ be the boundary states corresponding to the D-branes located at $x_0$
with $\eta$ appearing in the definition of the fermion boundary states and 
$\ket{B_1, \eta'}$ be an analogous states with different values of $x_1$ and $\eta'$. 
Let $\bra{B_1, \eta'}$ be the CPT conjugate states for $\ket{B_1, \eta'}$.
The evaluation of 
the overlap of the boundary state leads to 
\bea
Z_C&=& \int_0^\infty d\ell \bra{B_1, \eta'}e^{-2\pi\ell X^+H_C}\ket{B_2, \eta}   \nonumber \\
 &=& \int d\ell \frac{(1-\eta\eta'e^{-4\pi\ell\omega_0})^2
(1-\eta\eta'\prod_{n=1}^\infty e^{-4\pi\ell\omega_n})^4
(1-\eta\eta'\prod_{n=1}^\infty e^{-4\pi\ell\omega_n'})^4}
{(1-e^{-4\pi\ell\omega_0})^2
(1-\prod_{n=1}^\infty e^{-4\pi\ell\omega_n})^4(1-\prod_{n=1}^\infty 
e^{-4\pi\ell\omega_n'})^4} \nonumber \\
 & & \exp(-\frac{\hat{\omega_0}}{2\ap}((\vec{x_0^D})^2+(\vec{x_1^D})^2))
 \exp{-(\frac{\hat{\omega_0}}{2\ap}\frac{(e^{2\ell\hat{\omega_0}\pi}
+e^{-2\ell\hat{\omega_0}\pi})
((\vec{x_1^D})^2+(\vec{x_0^D})^2)-4\vec{x_1^D}\cdot\vec{x_0^D}}{e^{2\ell\hat{\omega_0}\pi}
-e^{-2\ell\hat{\omega_0}\pi}})}  \nonumber  \\
& & \exp(-\frac{\hat{\omega'_0}}{2\ap}((\vec{x_0^{'D}})^2+(\vec{x_1^{'D}})^2))
 \exp{-(\frac{\hat{\omega'_0}}{2\ap}\frac{(e^{2\ell\hat{\omega'_0}\pi}
+e^{-2\ell\hat{\omega'_0}\pi})
((\vec{x_1^{'D}})^2+(\vec{x_0^{'D}})^2)
-4\vec{x_1^{'D}}\cdot\vec{x_0^{'D}}}{e^{2\ell\hat{\omega'_0}\pi}
-e^{-2\ell\hat{\omega'_0}\pi}})}.  \nonumber
\eea
For the $DpDp$ boundary condition we have $\eta\eta'=1$ and $DpD\bar{p}$ boundary condition 
we have  $\eta\eta'=-1$. In the process of the calculation we crucially use the fact 
$\gamma^4\Omega\gamma^4\Omega=-1$. Since $\Omega$ commutes with $\gamma^{12349}$, 
two supermultiplets 
can be separately dealt with. One can check that the above result is coincident with the open 
string results up to overall normalization factor which can be absorbed into the normalization 
factor of the boundary state. This provides a consistency check for the boundary state construction 
carried out here.

\vs{.5cm}
\noindent
{\large\bf Acknowledgments}\\[.2cm]
We greatly appreciate Y. Michishita for explaining his results 
on the Type IIB boundary states to us. The work of J. P. is supported by the 
Korea Research Foundation (KRF) Grant KRF-2002-070-C00022. 


\newcommand{\J}[4]{{\sl #1} {\bf #2} (#3) #4}
\newcommand{\andJ}[3]{{\bf #1} (#2) #3}
\newcommand{\AP}{Ann.\ Phys.\ (N.Y.)}
\newcommand{\MPL}{Mod.\ Phys.\ Lett.}
\newcommand{\NP}{Nucl.\ Phys.}
\newcommand{\PL}{Phys.\ Lett.}
\newcommand{\PR}{Phys.\ Rev.}
\newcommand{\PRL}{Phys.\ Rev.\ Lett.}
\newcommand{\PTP}{Prog.\ Theor.\ Phys.}
\newcommand{\hepth}[1]{{\tt hep-th/#1}}



\begin{thebibliography}{99}

\bibitem{bfhp}
 M.\ Blau, J.\ Figueroa-O'Farrill, C.\ Hull and G.\ Papadopoulos,
 ``Penrose limits and maximal supersymmetry,''
 \hepth{0201081}, \J{Class.\ Quant.\ Grav.}{19}{2002}{L87};
 M.\ Blau, J.\ Figueroa-O'Farrill and G.\ Papadopoulos,
 ``Penrose limits, supergravity and brane dynamics,''
 \hepth{0202111} .

\bibitem{m}
 R.\ R.\ Metsaev, ``Type IIB Green Schwarz superstring in plane 
 wave Ramond Ramond background,''
 \hepth{0112044}, \J{\NP}{B625}{2002}{70}.

\bibitem{mt}
 R.\ R.\ Metsaev and A.\ A.\ Tseytlin, ``Exactly solvable model of 
 superstring in plane wave Ramond-Ramond background,''
 \hepth{0202109}.



\bibitem{bmn}
 D.\ Berenstein, J.\ Maldacena and H.\ Nastase, ``Strings in flat 
 space and pp waves from N=4 super Yang Mills,''
 \hepth{0202021}, \J{JHEP}{0204}{2002}{013}.


\bibitem{hyu074} S. Hyun and H. Shin, ``N=(4,4) Type IIA String Theory
  on PP-Wave Background,'' JHEP {\bf 0210} (2002) 070, hep-th/0208074.
  
\bibitem{sug029} K. Sugiyama and K. Yoshida, ``Type IIA String and
  Matrix String on PP-wave,'' Nucl. Phys. {\bf B644} (2002) 128,
  hep-th/0208029.
  
\bibitem{hyu158} S. Hyun and H. Shin, ``Solvable N=(4,4) Type IIA
  String Theory in Plane-Wave Background and D-Branes,''
  hep-th/0210158.

\bibitem{hyu090} S. Hyun and H. Shin, ``Branes from Matrix Theory in
  PP-Wave Background,'' Phys. Lett. {\bf B543} (2002) 115,
  hep-th/0206090.

\bibitem{hyunnew} S. Hyun and J. Park and S. Yi, ``Thermodynamic behavior 
 of  IIA string theory on a pp wave,'' hep-th/0304239. 
 
\bibitem{hyun123} S. Hyun, J. Park and H. Shin, ``Covariant Description 
of D-branes in IIA plane wave background,'' hep-th/0212343. 


\bibitem{gaber}
 O.\ Bergman, M.\ R.\ Gaberdiel and M.\ B.\ Green, ``D-brane 
 interactions in type IIB plane-wave background,''
 \hepth{0205183}.

\bibitem{gaber2} M. R. Gaberdiel and M. B. Green, ``The D-instanton
  and other supersymmetric D-branes in IIB plane-wave string theory,''
  hep-th/0211122.

\bibitem{gaber3} M. R. Gaberdiel, M. B. Green, S. Schafer-Nameki and A. Sinha, 
``Oblique and curved D-branes in IIB plane-wave string theory,'' hep-th/0306056.  

\bibitem{pol} J. Polchinski and Y. Cai, ``Consistency of open superstring 
theories,'' Nucl. Phy. {\bf B296} (1988) 91.


\bibitem{cal} C. Callan, C. Lovelace, C. Nappi and S. Yost, ``Adding holes 
and crosscaps to the superstring,'' Nucl. phy. {\bf 293} (1987) 83. 

\bibitem{rec} A. Recknagel and V. Schomerus, ``D-branes in Gepner models,''
hep-th/9712186.

\bibitem{rec2}A. Recknagel and V. Schomerus, ``Boundary deformation theory
and moduli spaces of D-branes,'' hep-th/9811237. 


\bibitem{shin} H. Shin, K. Sugiyama and K. Yoshida, ``Partition function 
and open/closed string duality in type IIA string theory on a pp-wave,''
hep-th/0306087.


\bibitem{yoji} Y. Michishita, unpublished.

\bibitem{sken} K. Skenderis and M. Taylor, ``Open strings in the plane wave 
background II: Superalgebras and spectra,'' hep-th/0212184.

\bibitem{sken2} K. Skenderis and M. Taylor, ``Branes in AdS and pp-wavw spacetimes,''
JHEP 0206 (2002) 025, hep-th/0204054; D. Freedman, K. Skenderis and M. Taylor, ``Worldvolume 
supersymmetries for branes in plane waves,'' hep-th/0306946.



\bibitem{peeters} P. Bain, K. Peeters and M. Zamalkar, ``D-branes in a plane 
wave from covariant open strings,'' Phy.rev. D67 (2003) 066001, hep-th/0208038.

\bibitem{cot} F. Bigazzi and A. Cotrone, ``On zero point energy, stability and Hagedorn 
behavior of Type IIB strings on pp-wave,'' hep-th/0306102.

\bibitem{park} Y. Michishita and J. Park, in preparation. 


  



\end{thebibliography}
\end{document}